\documentclass[10pt,final,doublecolumn]{IEEEtran}
\usepackage{amsmath}
\usepackage{latexsym}
\usepackage{graphicx}
\usepackage{bbding}
\usepackage{indentfirst}
\usepackage{algorithm,algorithmic}
\usepackage{setspace}
\usepackage{float}
\usepackage{epstopdf}
\usepackage{amssymb}
\usepackage{amsfonts}
\usepackage{enumerate}
\usepackage{multicol}
\usepackage{color}
\usepackage{slashbox}
\usepackage{bm}
\usepackage{amssymb}
\usepackage{stfloats}
\allowdisplaybreaks[4]
\begin{document}
\IEEEoverridecommandlockouts
\title{Robust Design for NOMA-based Multi-Beam LEO Satellite Internet of Things}
\author{Jianhang Chu, Xiaoming Chen, Caijun Zhong, and Zhaoyang Zhang
\thanks{Jianhang Chu ({\tt chujh@zju.edu.cn}), Xiaoming Chen ({\tt chen\_xiaoming@zju.edu.cn}), Caijun Zhong ({\tt caijunzhong@zju.edu.cn}) and Zhaoyang Zhang ({\tt ning\_ming@zju.edu.cn}) are with the College of Information Science and Electronic Engineering, Zhejiang University, Hangzhou 310027, China.}}\maketitle

\begin{abstract}
In this paper, we investigate the issue of massive access in a beyond fifth-generation (B5G) multi-beam low earth orbit (LEO) satellite internet of things (IoT) network in the presence of channel phase uncertainty due to channel state information (CSI) conveyance from the devices to the satellite via the gateway. Rather than time division multiple access (TDMA) or frequency division multiple access (FDMA) with multi-color pattern, a new non-orthogonal multiple access (NOMA) scheme is adopted to support massive IoT distributed over a very wide range.  Considering the limited energy on the LEO satellite, two robust beamforming algorithms against channel phase uncertainty are proposed for minimizing the total power consumption in the scenarios of noncritical IoT applications and critical IoT applications, respectively. Both thoeretical analysis and simulation results validate the effectiveness and robustness of the proposed algorithms for supporting massive access in satellite IoT.
\end{abstract}

\begin{IEEEkeywords}
B5G, NOMA, satellite IoT, multi-beam LEO satellite, massive access, robust design.
\end{IEEEkeywords}

\section{Introduction}
The Internet of Things (IoT) is changing various social and economic fields, e.g., industry, medicine, traffic, and agriculture \cite{IoT1,IoT2}. In order to unlock the potential of IoT, devices have to be interconnected wirelessly. Traditional IoT networks based on WiFi, narrowband IoT (NB-IoT) and LoRa access techniques are competent enough to fulfill communication tasks in cities and regular workplaces \cite{IoT3}. However, for some underpopulated special areas such as deserts, oceans and forests, it is impossible to construct tradition IoT networks owing to the geographical conditions and economic costs. Thus, satellite IoT is deemed to be a promising approach to compensate the shortage of terrestrial IoT \cite{SIoT1}. Actually, some companies and organizations such as Space X have launched many test satellites to realize global Internet service. In particular, the low-earth orbit (LEO) satellite becomes a better choice of an access point of satellite IoT based on B5G wireless networks due to its appealing characteristics, e.g., low power consumption and low transmission delay \cite{LEO1,LEO2}.

In order to serve multiple IoT devices distributed over a very large range, multi-beam technique is usually employed at the LEO satellite \cite{Multibeam1}. Specifically, the LEO satellite simultaneously generates multiple spot beams by a feed reflector antenna, and each beam covers a specified region \cite{Multibeam2}. By increasing the number of spots, it is possible to support the access of many devices. However, multi-beam communications lead to co-channel interference degrading the access performance. Traditionally, the multi-beam satellite system adopts frequency division multiplexing access (FDMA) such as four-color frequency reuse \cite{Fourcolor} to mitigate interference among adjacent beams. However, such a pattern can not effectively make full use of spectrum and thus full frequency reuse pattern with powerful precoding techniques was proposed in \cite{channel1}. Yet, full frequency reuse induces another major problem, namely, the large spectrum demands of feeder link (FL). In \cite{on board}, the authors proposed an on-board beamforming scheme to reduce the bandwidth of FL. Similarly, beamforming for multi-gateway multi-beam satellite systems \cite{multigateway} was also seemed to be an effective way to tackle the bandwidth problem. In general, these two approaches still need large bandwidth of FL which is several times larger than the bandwidth of beams and increase the cost of satellite system. Time division multiplexing access (TDMA) can avoid the defects as mentioned above. But in the application of massive IoT, TDMA may cause a high scheduling delay. Except for TDMA and FDMA, code-division multiple access (CDMA) has also been widely employed in satellite communications. There are a large amount of studies dedicating to compare the performance of CDMA or TDMA/FDMA in LEO mobile satellite communications \cite{CDMA1}-\cite{CDMA3}. In \cite{CDMA1}, the authors proved that a CDMA system had a greater capacity than an FDMA one without considering the effects of imperfect power control and the adjacent cells interference. When considering these factors, there was an opposite conclusion that TDMA/FDMA was superior to CDMA \cite{CDMA2}. Furthermore, the authors in \cite{CDMA3} estimated the capacity of a CDMA-based LEO satellite system with a comprehensive account of multiple access interference (MAI) and realistic evaluation of interference factors, and concluded that the capacity of CDMA was superior to that of FDMA/TDMA for a small power control error or a low SINR threshold. However, the feasibility of CDMA on LEO satellite is based on the scenario of mobile satellite applications which can use voice activity technique and other interference mitigation techniques to improve the capacity. In satellite IoT networks, due to the effect of MAI on the CDMA system capacity \cite{MAI1}, the massive access of IoT devices inevitably causes the serious MAI which leads to the sharp degradation of system performance. Therefore, traditional orthogonal multiple access schemes are not applicable to LEO multi-beam satellite IoT, which means that it is necessary to adopt non-orthogonal multiple access (NOMA) schemes \cite{NOMA1}, \cite{NOMA2}. Based on the NOMA schemes, it is likely to support massive IoT devices with limited spot beams. Besides, the power consumption of satellite can be reduced due to the decrease of radio frequency (RF) chains number. In \cite{satnoma}, the authors showed the feasibility of applying power-domain NOMA (PD-NOMA) in satellite IoT networks based on the characteristic of satellite channels. Combining with power allocation, PD-NOMA scheme can conquer the near-far effects in each beam which is a big problem in CDMA system. Moreover, the spectral efficiency of PD-NOMA is much higher than that of FDMA/TDMA/CDMA, which is a big advantage of the NOMA-based satellite IoT networks. However, the use of NOMA causes serious co-channel interference, especially in the context of massive IoT \cite{NOMA3}. Hence, it is essential to design spot beams to effectively eliminate the co-channel interference.

The design of spot beams requires accurate channel state information (CSI) at the LEO satellite. In general, the LEO satellite obtains CSI via the aid of the gateway (GW). Due to the round-trip delay and device mobility, it is difficult to acquire real-time CSI at the GW. In other words, channel uncertainty exists in the LEO satellite channels. As a result, the real performance of beamforming schemes would deteriorate if the LEO satellite designs spot beams based on perfect CSI directly. Therefore, it is necessary to consider the robust design to against channel uncertainty. Generally, there are three different schemes to achieve robust design in satellite communication: 1) Worst-case design based on a deterministic uncertainty model \cite{robust1}; 2) Optimizing average performance based on an expectation constraint \cite{robust2}; 3) Optimizing the performance with a certain outage level based on outage probabilistic model \cite{robust3}. In this paper, we consider the latter two schemes to design the robust spot beams according to the characteristics of channel uncertainty in satellite communications.

\subsection{Previous Works}
As a promising technique to improve network capacity and spectrum efficiency, NOMA has been applied for terrestrial IoT network \cite{terIoT1}, \cite{terIoT2}. For satellite IoT, the authors in \cite{summary} summarized the state of the art in satellite communication (SatCom) and the most promising applications in SatCom such as satellite IoT, and the authors in \cite{LEO2} analyzed the details of a LEO satellite IoT network. Besides, the advantages and implementation details of satellite communication based on NOMA was also well studied in \cite{NOMA-SAT1}-\cite{NOMA-SAT5}. To be specific, \cite{NOMA-SAT1} elaborately provided different non-orthogonal schemes that are suitable for the forward link and \cite{NOMA-SAT5} studied the application of power-domain NOMA scheme in SatCom systems. The authors in \cite{NOMA-SAT2} addressed the quality-of-service (QoS) guaranteed resource allocation problem in NOMA-based satellite industrial IoT networks. \cite{NOMA-SAT3} extended the cooperative NOMA with asynchronous channel in downlink of multibeam satellite networks. Also, \cite{NOMA-SAT4} utilized multiuser cooperative scheme to conduct inter-usr interference cancellation and analyzed the performance of secrecy rate in the frequency-domain NOMA system.

However, in the aforementioned papers, the analysis of NOMA-based satellite system is based on perfect CSI which is unpractical in the real SatCom environment. In general, due to delay and error during the CSI conveyance from the devices to the satellite via the gateway, there exists channel phase error for the CSI at the satellite. Thus, it is necessary to consider the impact of imperfect CSI on NOMA-based satellite IoT system. Indeed, there have some researches dedicated to multi-beam SatCom in the presence of channel uncertainty \cite{robust1}-\cite{robust3}, \cite{ro-SAT1}-\cite{channel2}. Similar to channel models of the terrestrial, \cite{robust1} and \cite{ro-SAT1} assumed that channel uncertainty is the additive norm-bounded estimation error and separately proposed a robust beamforming design for maximizing the sum secrecy rate in multibeam SatCom systems. Yet, it is more realistic to consider the multiplicative phase error in satellite channels due to the special characteristics in satellite channels \cite{phase}. \cite{robust2} and \cite{ro-SAT2} both focused on the power minimization problem with expectation constraint of user's SINR. Moreover, the authors in \cite{robust3} and \cite{channel2} investigated the robust design for multigroup multicast precoding for multibeam SatCom systems with full frequency reuse. By considering the outage probability constraint of each user' SINR, \cite{channel2} adopted the central limit theorem to transform the non-convex outage probability constraint and change the original problem into a convex one which can be solved by standard convex solvers. In addition, the above researches tackled the rank-one constraint of beamforming matrixes with classic Gaussian Randomization solution \cite{rank}.

\subsection{Motivations and Contributions}
This paper dedicates to investigate the robust beamforming design for a NOMA-based satellite IoT network. To the best of our knowledge, the total power minimization problem in NOMA-based satellite IoT network with multiplicative channel phase error is still a new yet challenging domain in NOMA-based satellite IoT related fields. Thus, these observations motivate us to make efforts on this research. The main contributions of this paper are summarized as follows:
\begin{enumerate}
\item We present a NOMA-based LEO multi-beam satellite IoT network to make up the defects of terrestrial IoT network. Specifically, the proposed LEO multi-beam satellite IoT network can provide a wide coverage for a massive number of IoT devices with low power.

\item We design a practical LEO multi-beam satellite framework for massive IoT, where the GW only acquires partial CSI. Considering two typical IoT application scenarios, we separatively formulate the robust beamforming design for minimizing the total transmit power with satisfying different SINR constraints of users and per-antenna power constraint.

\item For tackling the non-convexity of two robust designs, we utilize a series of mathematical tools to reformulate original problems into the approximately equivalent convex ones. We proposed two novel iteratively algorithms combining with a penalty function rather than the classic Gaussian Randomization scheme to solve the convex problems. Furthermore, we investigate the impact of imperfect successive interference cancellation (SIC) and on the performance of the LEO multi-beam satellite IoT network. Simulation results show the effectiveness and robustness of the two proposed algorithms.
\end{enumerate}

The rest of this paper is organized as follows: Section \uppercase\expandafter{\romannumeral2} introduces a multi-beam LEO satellite IoT network. Section \uppercase\expandafter{\romannumeral3} concentrates on the design of two robust algorithms based on different IoT application scenarios to minimize the total power consumption on the LEO satellite. Section \uppercase\expandafter{\romannumeral4} provides numerical results to evaluate the effectiveness and robustness of the proposed algorithms. Finally, we conclude the paper in Section \uppercase\expandafter{\romannumeral5}.

$Notation:$ We use bold lowercase and uppercase letters to denote column vectors and matrices, $\mathbb{R}^{m\times n}$, $\mathbb{H}^{m\times n}$ and $\mathbb{R}^{m}$ denote $m\times n$ real and complex matrices, $m$-dimensional real vector, respectively. $\mathcal{S}^K$ and $\mathcal{K}^K$ denote the symmetric and skew-symmetric matrixs. $(\cdot)^H$ and $(\cdot)^T$ denote Hermitian transpose and transpose, $\|\cdot\|$ and $|\cdot|$ denote Euclidean norm and absolute value, $\text{tr}(\cdot)$ and $\text{Rank}(\cdot)$ denote trace and rank of a matrix, $\odot$ to denote Hadamard product. We use diag($\textbf{x}$) to denote the diagonal matrix with the elements of main diagonal constituted by $\textbf{x}$, $[\textbf{X}]_{m,n}$ to denote the $[m,n]$th element of $\textbf{X}$.

\section{System Model}
\begin{figure}[t] \centering
\includegraphics [width=0.45\textwidth] {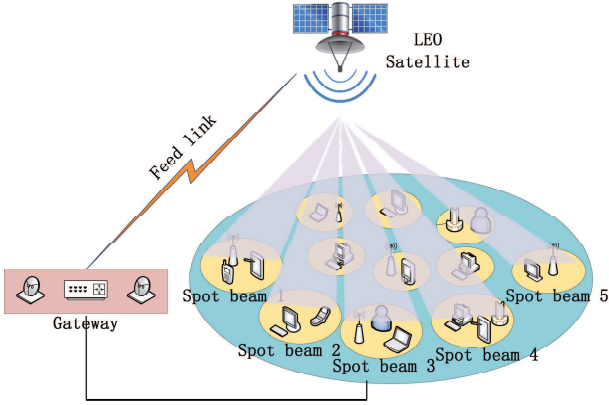}
\caption {A model of B5G multi-beam LEO satellite IoT network.}
\label{Fig1}
\end{figure}

In this section, we consider a B5G multi-beam LEO satellite IoT network, where a satellite equipped with an array fed reflector antenna communicates with $N$ single-antenna IoT user equipments (UEs). As shown in Fig. 1, IoT UEs obtain channel state information (CSI) about downlink channels by channel estimation and send it to a GW, which conveys the CSI to the satellite through  a high-capacity feedback link\footnote{In this paper, we assume the feedback link is error-free. In practice, there may exist error, resulting in imperfect CSI. Imperfect CSI in satellite communications is usually modeled as channel phase error as analyzed later.}. Then, the satellite utilizes the array fed reflector antenna to generate $M$ spot beams according to the obtained CSI and broadcasts the beamformed signals over the downlink channels. The array fed reflector antenna comprises a feed array with $K$ feeds, a beamforming network, and a reflector antenna. In general, the beamforming network has two kinds of structures, namely, single feed per beam (SFB) with $M = K$ and multiple feeds per beam (MFB) with $M < K$. For convenience of flexibly controlling the shape and the number of beams, we adopt the MFB structure in the considered LEO satellite to generate $M$ beams in this paper. Different from traditional multicast communications with a multi-beam LEO satellite, NOMA techniques are applied to support massive IoT distributed in a large range over a limited radio spectrum. Specifically, each spot beam covers a specific region. The UEs in a region share the same beam but decode different data streams, and thus it is possible to realize massive access with a small number of RF chains. Without loss of generality, we assume that the satellite coverage area is split into $M$ regions, and the $m^{th}$ region contains $N_m$ active IoT UEs. For ease of analysis, we use UE$_{m,n}$ to denote the $n^{th}$ UE in the $m^{th}$ region in the rest of paper. In what follows, we introduce the massive access scheme for B5G multi-beam LEO satellite IoT.
\subsection{Channel Model}
According to the signal propagation characteristics of LEO satellite communications, the downlink channel between the satellite and the UE$_{m,n}$ can be expressed as \cite{channel1}, \cite{channel2}
\begin{equation}
\textbf{h}_{m,n}=\sqrt{C_{m,n}}\textbf{b}_{m,n}^{1/2}\odot\textbf{r}_{m,n}^{1/2}\odot \exp{\left\{j\boldsymbol{\theta}_{m,n}\right\}}, \label{eqn1}
\end{equation}
where $C_{m,n}$ is the large-scale fading efficient, which is given by
\begin{equation}
C_{m,n}=(\frac{\upsilon} {4\pi fd_{0}})^2 \frac{G_{m,n}}{\kappa BT}, \label{eqn2}
\end{equation}
where $(\frac{\upsilon} {4\pi fd_{0}})^2$ is the free space loss (FPL) with $\upsilon$ being the light speed, $f$ being the carrier frequency and $d_0$ being propagation distance, $G_{m,n}$ is the receive antenna gain of the UE$_{m,n}$, $\kappa$ is the Boltzman's constant, $B$ is the carrier bandwidth, and $T$ is the receive noise temperature. $\textbf{b}_{m,n}$ is a $K$-dimensional beam radiation pattern vector, which of the $k^{th}$ element, namely the beam gain from the $k^{th}$ feed to the UE$_{m,n}$, can be approximated as
\begin{equation}
\textbf{b}_{m,n}(k)=G_m\left(\frac{J_1(u_k)}{2u_k}+36\frac{J_3(u_k)}{u_k^3}\right)^2, \label{eqn3}
\end{equation}
where $G_m$ represents the maximum satellite antenna gain for the $m^{th}$ beam and $u_k=2.07123\frac{\sin(\varphi_{m,n})}{\sin(\varphi_{m,3\text{dB}})}$. Here, $\varphi_{m,n}$ is the angle between the $k^{th}$ feed and the UE$_{m,n}$, and $\varphi_{m,\text{3dB}}$ is a constant which is equal to the 3 dB angle for the $m^{th}$ beam. $J_1$ and $J_3$ are the first and third order of first-kind Bessel function, respectively. Furthermore, $\textbf{r}_{m,n}$ is a $K$-dimensional rain attenuation coefficient vector. In the form of dB, $\textbf{r}_{m,n}^{dB}(k)=20\log_{10} \textbf{r}_{m,n}(k)$, commonly follows lognormal random distribution $\ln(\textbf{r}_{m,n}^{dB}(k))\sim \mathcal{CN}(\mu_r,\sigma_r^2)$ \cite{rain}. Finally, $\boldsymbol{\theta}_{m,n}$ is a $K$-dimensional channel phase vector with each element independently obeying uniformly distributed between 0 and 2$\pi$.

Apparently, the amplitude of the satellite channel is determined by the beam gain, rain attenuation and large scale fading factor, which can be considered as a constant over the intervals of interest. However, the phase of each channel element is influenced by multiple time varying factors. For example, fog, rain and atmospheric absorption can lead to serious time varying phase variations on channel elements, and these variations change faster than those for amplitude. As a result, there may exist channel phase estimation error at the IoT UEs. In general, the relationship between the actual channel phase vector $\boldsymbol{\theta}_{m,n}$ and the estimated channel phase vector $\bar{\boldsymbol{\theta}}_{m,n}$ can be expressed as
\begin{equation}
\boldsymbol{\theta}_{m,n}=\overline{\boldsymbol{\theta}}_{m,n}+\textbf{e}_{m,n}, \label{eqn4}
\end{equation}
where $\textbf{e}_{m,n}$ is a $K$-dimensional channel phase error vector with independently and identical distributed (i.i.d.) Gaussian random element, namely, $\textbf{e}_{m,n}\sim \mathcal{N}(\mathbf{0},\sigma^2_{m,n}\mathbf{C}_{m,n})$ with $\sigma^2_{m,n}$ being the variance of the phase error and $\mathbf{C}_{m,n}$ being the normalized covariance matric. Then, the relationship between the real CSI $\textbf{h}_{m,n}$ and the obtained CSI $\bar{\textbf{h}}_{m,n}$ can be modeled as
\begin{equation}
\textbf{h}_{m,n}=\overline{\textbf{h}}_{m,n}\odot\textbf{q}_{m,n}=\text{diag}(\overline{\textbf{h}}_{m,n})\textbf{q}_{m,n}, \label{eqn5}
\end{equation}
where $\textbf{q}_{m,n}\triangleq\ \exp{\left\{j\textbf{e}_{m,n}\right\}}$.
\subsection{Signal Model}
As mentioned earlier, in order to support massive access with a finite number of beams, the NOMA technique is adopted in the satellite IoT. Based on the obtained CSI, the satellite carries out superposition coding before broadcasting signals to the UEs. First, the satellite constructs the transmit signal $x_m$ for the $m^{th}$ region, $\forall m$, as follows:
\begin{equation}
x_m=\sum\limits_{n=1}^{N_m}\sqrt{\alpha_{m,n}}s_{m,n},\label{eqn6}
\end{equation}
where $s_{m,n}$ is the Gaussian distributed signal of unit norm for the UE$_{m,n}$, and $\alpha_{m,n}$ is the intra-region power allocation factor, which can coordinate the intra-region interference, and satisfies the following constraint:
\begin{equation}
\sum\limits_{n=1}^{N_m}\alpha_{m,n}\leq 1,\forall m, \label{eqn7}
\end{equation}
Then, it constructs the total transmit signal $\textbf{x}$ as below:
\begin{equation}
\textbf{x}=\sum\limits_{m=1}^{M}\textbf{w}_mx_m,\label{eqn8}
\end{equation}
where $\textbf{w}_m$ is a $K$-dimensional beamforming vector designed for the $m^{th}$ region based on available CSI to reduce the inter-region interference. Finally, the satellite broadcasts the superposition coded signal $\mathbf{x}$ to the UEs. Hence, the received signal at the UE$_{m,n}$ is given by
\begin{eqnarray}
y_{m,n}=\textbf{h}_{m,n}^H\textbf{x}+n_{m,n},\label{eqn9}
\end{eqnarray}
where $n_{m,n}$ is additive white Gaussian noise (AWGN) at the UE$_{m,n}$ with variance $\sigma^2_0$.

Owing to the effect of superposition coding, there exists intra-region interference among UEs in the same ragion. Therefore, the UEs carry out successive interference cancellation (SIC) to decrease intra-region interference. Based on the channel quality indicators from the IoT UEs, the satellite can line the effective channel gains up in each region and conveys it to the UEs via downlink channels. Without loss of generality, we presume that the effective channel gains of the $m^{th}$ region have a descending order as follows:
\begin{equation}
|\mathbf{h}_{m,1}^H\mathbf{w}_m|^2\geq|\mathbf{h}_{m,2}^H\mathbf{w}_m|^2\geq\cdots\geq|\mathbf{h}_{m,N_m}^H\mathbf{w}_m|^2.\label{eqn10}
\end{equation}
According to the principle of SIC, the $i^{th}$ UE decodes the interfering signals and removes it based on the reverse order in Eq. (\ref{eqn10}), and finally demodulates its desired signal. However, due to the hardware defects of IoT UEs, the consistent change of circumstance in satellite-ground links and other factors, decoding errors of the interfering signal from the UEs with weak channel gains may occur, resulting in residual interference after SIC, namely imperfect SIC. According to a linear model of imperfect SIC \cite{11}, the post-SIC signal at the UE$_{m,n}$ can be written as
\begin{eqnarray}
y_{m,n}\!\!\!\!&=&\!\!\!\!{\underbrace{\textbf{h}_{m,n}^H\textbf{w}_{m}\sqrt{\alpha_{m,n}}s_{m,n}}_{\textrm{Desired signal}}}\nonumber\\
&&\!\!\!\!+{\underbrace{\textbf{h}_{m,n}^H\textbf{w}_{m}\sum\limits_{i=1}^{n-1}\sqrt{\alpha_{m,i}}s_{m,i}}_{\textrm{Intra-region interference}}}\nonumber\\
&&\!\!\!\!+{\underbrace{\textbf{h}_{m,n}^H\textbf{w}_{m}\sum\limits_{i=n+1}^{N_m}\sqrt{\eta_{m,n} \alpha_{m,i}}s_{m,i}}_{\textrm{Residual intra-region interference}}}\nonumber\\
&&\!\!\!\!+{\underbrace{\sum\limits_{j=1,j\neq m}^{M}\textbf{h}_{m,n}^H\textbf{w}_{j}\sum\limits_{i=1}^{N_m}\sqrt{\alpha_{j,i}}s_{j,i}}_{\textrm{Inter-region interference}}}+{\underbrace{n_{m,n}}_{\mathrm{AWGN}}},\nonumber\\\label{eqn11}
\end{eqnarray}
where $\eta_{m,n} \in$ [0,1] represents the coefficient of imperfect SIC associated with the UE$_{m,n}$, which can be acquired by long-term measurement. Note that different values of $\eta_{m,n}$ mean different functions of SIC. Firstly, $\eta_{m,n} = 0$ means that the UE$_{m,n}$ can carry out SIC perfect to cancel the intra-region interference from weaker UEs. Secondly, $0 <\eta_{m,n} < 1$ represents the situation that the UE$_{m,n}$ is able to conduct SIC, but SIC is imperfect. Thirdly, $\eta_{m,n} = 1$ indicates that the UE$_{m,n}$ has no capability of performing SIC. Thus, the signal-to-interference-plus-noise ratio (SINR) at the UE$_{m,n}$ can be expressed as
\begin{equation}
\Gamma_{m,n}=\frac{\alpha_{m,n}|\mathbf{h}_{m,n}^H\mathbf{w}_m|^2}{\sum\limits_{j=1}^{M}\sum\limits_{i=1}^{N_m}\beta_{j,i}^{m,n}\alpha_{j,i}|\mathbf{h}_{m,n}^H\mathbf{w}_j|^2+\sigma^2_0},\label{eqn12}
\end{equation}
where $\beta_{j,i}^{m,n}$ is an instrumental variable which is defined as
\begin{equation}
\beta_{j,i}^{m,n}=
\begin{cases}
0, & \mbox{if }j=m\mbox{ and }i=n,\\
\eta_{m,n}, & \mbox{if }j=m\mbox{ and }i>n,\\
1, & \text{otherwise}.
\end{cases}
\end{equation}\label{eqn13}

It is seen from (\ref{eqn12}) that the performance of the LEO satellite IoT network is closely related to the beam $\mathbf{w}_m, \forall m$. In particular, the beams determine the gains of both the desired signal and the interfering signals. Therefore, it is necessary to design spot beams from the perspective of improving the overall performance. Considering the limited energy at the practical LEO satellite, we dedicate to design beams to minimize the total power consumption in the next section.

\section{Robust Design of B5G Multi-Beam LEO Satellite IoT}
In this section, we design robust beamforming for multi-beam LEO satellite IoT networks in the presence of channel phase error for minimizing the total transmit power consumption\footnote{In general, the total power consumption includes the signal transmit power, the constant circuit power consumption per antenna relating with transmit filter, mixer, frequency synthesizer, and digital-to-analog converter and the basic power consumed at the satellite. Since the circuit power and the basic power are independent of the signal transmit power, we consider the minimization of only the signal transmit power.} based on the characteristics of IoT applications, i.e., noncritical applications and critical applications.

\subsection{Noncritical Robust Design}
First, we consider the scenario of noncritical IoT applications, e.g., agriculture, entertainment, and home. For these noncritical IoT applications, the long-term received signal quality is more important than the instantaneous one. In this case, the robust design can be formulated as the following optimization problem:
\begin{subequations}\label{OP0}
\begin{equation}
\min\limits_{\boldsymbol{\mathbf{w}}_m,\forall m}\sum_{m=1}^{M}\left\|\textbf{w}_m\right\|^2
\end{equation}
\begin{equation}
\textrm{s.t.}\:\mathbb{E}\left\{\Gamma_{m,n}\right\}\geq\gamma_{m,n},\forall m,n, \label{c11}
\end{equation}
\begin{equation}
\left[ \sum_{m=1}^{M}\textbf{w}_m\textbf{w}_m^H \right]_{k,k}\leq P_{k},\forall k,
\end{equation}
\end{subequations}
where $\gamma_{m,n} > 0$ is the required minimum average SINR for the UE$_{m,n}$ and $P_{k}$ is the power constraint of the $k^{th}$ feed. It is obvious that problem (\ref{OP0}) is non-convex. To this end, we introduce an auxiliary variable $\textbf{W}_m = \textbf{w}_m\textbf{w}_m^H$ and reformulate problem (\ref{OP0}) as
\begin{subequations} \label{OP1}
\begin{equation}
\min\limits_{\boldsymbol{\mathbf{W}}_m, \forall m}\sum_{m=1}^{M}\text{tr}(\textbf{W}_m),
\end{equation}
\begin{equation}
\textrm{s.t.}\:\mathbb{E}\left\{\Gamma_{m,n}\right\}\geq\gamma_{m,n},\forall m,n, \label{op11}
\end{equation}
\begin{equation}
\left[ \sum_{m=1}^{M}\textbf{W}_m \right]_{k,k}\leq P_{k},\forall k,\label{op12}
\end{equation}
\begin{equation}
\textbf{W}_m\succeq 0,\forall m,\label{op13}
\end{equation}
\begin{equation}
\text{Rank}(\textbf{W}_m)=1,\forall m,\label{op14}
\end{equation}
\end{subequations}
To solve problem (\ref{OP1}), firstly, we take a few transformation and approximation on the constraint (\ref{op11}), which can be rewritten as
\begin{equation}
\Gamma^{'}_{m,n}=\frac{\alpha_{m,n}\text{tr}(\textbf{H}_{m,n}\textbf{W}_{m})}{t_1\text{tr}(\textbf{H}_{m,n}\textbf{W}_{m})+\sum\limits_{j=1,j\neq m}^{M}t_2\text{tr}(\textbf{H}_{m,n}\textbf{W}_{j})+\sigma^2_0} \label{eqn15}
\end{equation}
where $t_1=\sum_{i=1}^{n-1}\alpha_{m,i}+\sum_{i=n+1}^{N_m}\eta_{m,n}\alpha_{m,j}$, $t_2=\sum_{i=1}^{N_m}\alpha_{j,i}$ and $\textbf{H}_{m,n} = \textbf{h}_{m,n} \textbf{h}_{m,n}^H$. As noted in Section \uppercase\expandafter{\romannumeral2}, $\textbf{H}_{m,n}$ can be expressed as
\begin{eqnarray}
\textbf{H}_{m,n}&=& \text{diag}(\overline{\textbf{h}}_{m,n})\textbf{q}_{m,n}\textbf{q}_{m,n}^H\text{diag}(\overline{\textbf{h}}_{m,n}^H)\nonumber\\\label{eqn14}
&=& \text{diag}(\overline{\textbf{h}}_{m,n})\textbf{Q}_{m,n}\text{diag}(\overline{\textbf{h}}_{m,n}^H). \label{eqn16}
\end{eqnarray}
Then, we set $\textbf{Q}^{'}_{m,n}=\mathbb{E}\left\{\textbf{Q}_{m,n}\right\}$, whose diagonal elements are all ones. Due to Gaussian distribution of $e_{m,n}$, the off-diagonal elements of $\textbf{Q}^{'}_{m,n}$ can be calculated as
\begin{eqnarray}
[\textbf{Q}^{'}_{m,n}]_{l,s}&=&\mathbb{E}\left\{\exp\left\{j[\textbf{e}_{m,n}]_l\right\}\right\}\mathbb{E}\left\{\exp\left\{-j[\textbf{e}_{m,n}]_s\right\}\right\}\nonumber\\
&=&\exp{\left\{-j\sigma^2_{m,n}\right\}}.\label{eqn17}
\end{eqnarray}
Therefore, the $(l,s)^{th}$ element of $\textbf{Q}^{'}_{m,n}$ takes the following value:
\begin{equation}
[\textbf{Q}^{'}_{m,n}]_{l,s}=
\begin{cases}
1, & \mbox{if }l=s\\
\exp{\left\{-j\sigma^2_{m,n}\right\}}, & \text{otherwise}.
\end{cases}  \label{eqn18}
\end{equation}
Hence, referring to \cite{expectation}, for nonnegative random variables $\text{X}$ and $\text{Y}$, we can have the following approximation:
\begin{equation}
\mathbb{E} \left\{\log_2\left(1+\frac{\text{X}}{\text{Y}}\right)\right\} \approx\log_2 \left(1+\frac{\mathbb{E}\left\{\text{X}\right\}}{\mathbb{E}\left\{\text{Y}\right\}} \right). \label{expec}
\end{equation}
According to (\ref{expec}), constraint (\ref{op11}) can be approximated by (\ref{eqn19}) at the top of the next page, where $\textbf{T}_{m,n}=t_1\textbf{W}_m+t_2\sum_{j=1,j\neq m}^{M}\textbf{W}_j$. To be a convex constraint condition, we transform (\ref{eqn19}) and get the following convex constraint (15$b^{'}$):
\newcounter{TempEqCnt}  
\setcounter{TempEqCnt}{21} 
\setcounter{equation}{20}
\begin{figure*}[ht]
\begin{eqnarray}
\mathbb{E}\left\{\Gamma^{'}_{m,n}\right\}\approx\frac{\alpha_{m,n}\mathbb{E}\left\{\text{tr}(\textbf{H}_{m,n}\textbf{W}_{m})\right\}}{t_1\mathbb{E}\left\{\text{tr}(\textbf{H}_{m,n}\textbf{W}_{m})\right\}+\sum\limits_{j=1,j\neq m}^{M}t_2\mathbb{E}\left\{\text{tr}(\textbf{H}_{m,n}\textbf{W}_{j})\right\}+\sigma^2_0}=\frac{\alpha_{m,n}\text{tr}(\text{diag}(\textbf{h}_{m,n})\textbf{Q}^{'}_{m,n}\text{diag}(\textbf{h}^H_{m,n})\textbf{W}_m)}{\text{tr}(\text{diag}(\textbf{h}_{m,n})\textbf{Q}^{'}_{m,n}\text{diag}(\textbf{h}^H_{m,n})\textbf{T}_{m,n})+\sigma^2_0}\geq\gamma_{m,n},\label{eqn19}
\end{eqnarray}
\hrulefill
\end{figure*}
\setcounter{equation}{\value{TempEqCnt}}
\begin{equation}
\text{tr}\left(\text{diag}(\textbf{h}_{m,n})\textbf{Q}^{'}_{m,n}\text{diag}(\textbf{h}^H_{m,n})\textbf{T}^{'}_{m,n}\right)-\gamma_{m,n}\sigma^2_0\geq0,\label{eqn20}
\end{equation}
where $\textbf{T}^{'}_{m,n}=\alpha_{m,n}\textbf{W}_m-\gamma_{m,n}\textbf{T}_{m,n}$.

However,  problem (\ref{OP1}) is still nonconvex owing to the rank-one constraint on $\textbf{W}_m,\forall m$. To guarantee a rank-one solution $\textbf{W}_m^*, \forall m$, we insert an iterative penalty function (IPF) into the objective function \cite{penalty}. Note that $\textbf{W}^*_m,\forall m$, is a positive semidefinite matrix which means that each eigenvalue $\lambda^*_m$ of $\textbf{W}^*_m$ satisfies $\lambda^*_m\geq0$. Intuitively, rank-one constraint implies that only one eigenvalue $\lambda^*_{m,\max}$ is larger than zero. Thus, according to the fact $\text{tr}(\textbf{W}_m)=\sum\limits_{i=1}^K\lambda_{m,i}$, we can replace the constraint (\ref{op14}) with the following equation:
\begin{equation}
\text{tr}\left(\textbf{W}_m\right)-\sum\limits_{i=1}^K\lambda_{m,\max}=0. \label{new1}
\end{equation}
Then, we can build a penalty function to improve the objective function of problem (\ref{OP1}) with the constraint (\ref{new1}):
\begin{equation}\label{newop1}
\min\limits_{\boldsymbol{\mathbf{W}}_m, \forall m}\sum_{m=1}^{M}\text{tr}\left(\textbf{W}_m\right)+\rho_1\sum\limits_{m=1}^M \left(\text{tr}(\textbf{W}_m)-\lambda_{m,\max}\right),
\end{equation}
where $\rho_1$ is the penalty factor. Unfortunately, the new objective function is non-convex due to the existence of the penalty function. To tackle this issue, we adopt an iterative method to acquire a convex one. To be specific, for the solution $\textbf{W}^{(t)}_m$ in the $t^{th}$ iteration, we have following inequality:
\begin{equation}
\begin{aligned}
&\text{tr}\left(\textbf{W}^{(t+1)}_m\right)-\left(\textbf{v}^{(t)}_{m,\max}\right)^H\textbf{W}^{(t+1)}_m\textbf{v}^{(t)}_{m,\max}\geq\\
&\text{tr}\left(\textbf{W}^{(t+1)}_m\right)-\lambda^{(t+1)}_{m,\max}\geq0. \label{new2}
\end{aligned}
\end{equation}
where $\textbf{v}_{m,\max}$ is the unit eigenvector corresponding to $\lambda_{m,\max}$. Then, problem (\ref{OP1}) can be reformulated as (\ref{OP2}) at the top of the next page. It is obvious that the value of the penalty factor will affect the convergence speed of the iteration-based objective function. Thus, it is necessary to choose a proper penalty factor at first. It is worth pointing out that due to the existence of a lower bound in (\ref{new2}), the objective function can be converged eventually. In other words, the rank-one constraint can be satisfied. Finally, problem (\ref{OP2}) is a convex problem, which can be effectively solved by some off-the-shelf optimization softwares, e.g., CVX. Afterward, we can obtain a suboptimal solution of the original problem (\ref{OP0}) as follows:
\newcounter{TempEqCnt1}  
\setcounter{TempEqCnt1}{26} 
\setcounter{equation}{25}
\begin{figure*}[ht]
\begin{equation}  \label{OP2}
\min\limits_{\boldsymbol{\mathbf{W}}_m, \forall m}\sum_{m=1}^{M}\text{tr}\left(\textbf{W}^{(t+1)}_m\right)+\rho_1\sum\limits_{m=1}^M\left(\text{tr}\left(\textbf{W}^{(t+1)}_m\right)-\left(\textbf{v}^{(t)}_{m,max}\right)^H\textbf{W}^{(t+1)}_m\textbf{v}^{(t)}_{m,max}\right)
\end{equation}
\begin{equation}
\textrm{s.t.}\:(15b^{'}),(\ref{op12}),(\ref{op13}). \nonumber
\end{equation}
\hrulefill
\end{figure*}
\setcounter{equation}{\value{TempEqCnt1}}
\begin{equation}\label{new3}
\textbf{w}^*_{m}=\sqrt{\lambda_{m,\max}}\textbf{v}_{m,\max}.
\end{equation}
As a result, the noncritical robust design can be summarized as Algorithm 1.
\begin{algorithm}
\caption{: Noncritical Robust Design of Multi-Deam LEO Satellite IoT for the Total Power Consumption Minimization}
\label{alg1}
\hspace*{0.02in} {\bf Input:}
$K,M,N_m,\alpha_{m,n},\gamma_{m,n},\sigma_0^2$, and $P_k$.\\
\hspace*{0.02in} {\bf Output:}
$\textbf{w}_m$.
\begin{algorithmic}[1]
\STATE{\textbf{Set} accuracy $\epsilon_1$, maximal iteration number $T_{\max}$, penalty factor $\rho_1$ and coefficient $\kappa_1$.}
\STATE{\textbf{Initialize} feasible solution  $\textbf{W}^{(0)}_m$ by solving problem (\ref{OP1}) dropping constraint (\ref{op14}).}
\STATE{\textbf{repeat}}
\STATE{\quad solve problem (\ref{OP2}) by CVX, then obtain $\textbf{W}_m^{(t)}$.}
\STATE{\quad\textbf{if} $\left|\text{tr}\left(\textbf{W}^{(t)}_m\right)-\lambda^{(t)}_{m,\max}\right|>\epsilon_1$ \textbf{then}}
\STATE{\qquad update the penalty factor $\rho^{(t)}_1=\kappa_1\rho^{(t+1)}_1$.}
\STATE{\quad\textbf{end if}}
\STATE{\quad Update iteration number $t=t+1$.}
\STATE{\textbf{until} $t=T_{\max}$ or solution converges.}
\STATE{\textbf{Finally}, use eigenvalue decomposition (EVD) to $\textbf{W}_m^{(t)}$ and obtain $\textbf{w}^*_{m}$ according to (\ref{new3}).}
\end{algorithmic}
\end{algorithm}
\subsection{Critical Robust Design}
Then, we consider the scenario of critical IoT applications, e.g., medicine, traffic, and industry. For these critical IoT applications, the instantaneous SINR has to satisfy a given condition. To guarantee the SINR performance over fading channels in the presence of channel phase error, we impose an outage probability constraint on the design of transmit beams. Thus, the critical robust design can be formulated as
\begin{subequations} \label{OP3}
\begin{equation}
\min\limits_{\boldsymbol{\mathbf{W}}_m, \forall m}\sum_{m=1}^{M}\text{tr}(\textbf{W}_m),
\end{equation}
\begin{equation}
\textrm{s.t.}\:\textbf{Pr}\left\{\Gamma_{m,n}\geq\gamma_{m,n}\right\}\geq1-p_{m,n},\forall m,n, \label{op31}
\end{equation}
\begin{equation}
\left[ \sum_{m=1}^{M}\textbf{W}_m \right]_{k,k}\leq P_{k},\forall k,\label{op32}
\end{equation}
\begin{equation}
\textbf{W}_m\succeq 0,\forall m,\label{op33}
\end{equation}
\begin{equation}
\text{Rank}(\textbf{W}_m)=1,\forall m,\label{op34}
\end{equation}
\end{subequations}
where $p_{m,n}$ is the SINR outage probability threshod for the UE$_{m,n}$. Unfortunately, problem (27) is also not convex. To convert problem (\ref{OP3}) into a convex problem, we should transform the probabilistic constraint into a series of convex conditions. In what follows, we introduce two useful lemmas for convexity transformation.

First, we need to take a pretreatment into the constraint (\ref{op31}), which can be expressed as
\begin{equation}
\text{Pr}\left\{\textbf{q}^H_{m,n}\textbf{Z}_{m,n}\textbf{q}_{m,n}-\sigma^2_0\geq0\right\}\geq1-p_{m,n},\forall m,n, \label{eqn21}
\end{equation}
where $\textbf{Z}_{m,n}=\left[\frac{\alpha_{m,n}}{\gamma_{m,n}}-t_1\right]\textbf{W}_m-\sum\limits_{j=1,j\neq m}^{M}t_2\textbf{W}_j$, $t_1$ and $t_2$ are the same as before. Then, we utilize the Taylor's expansion to approximate the term $\textbf{q}^H_{m,n}\textbf{Z}_{m,n}\textbf{q}_{m,n}$, which is shown in Lemma 1.

\emph{Lemma 1:} For Gaussian random vector $\boldsymbol{\theta}$, complex exponential Gaussian vector $\textbf{x}=(e^{j\theta_1},\cdot\cdot\cdot,e^{j\theta_K})$, and $K$-order Hermitian matrix $\textbf{Z}$ with symmetric real part $\textbf{A}\in\mathcal{S}^K$ and skew-symmetric imaginary part $\textbf{B}\in\mathcal{K}^K$, we can obtain the second-order Taylor's expansion expression of $\textbf{x}^H\textbf{Z}\textbf{x}$ as follows:
\begin{equation}
\textbf{x}^H\textbf{Z}\textbf{x}=\sum\limits_{i,j}\textbf{Z}_{i,j}+\boldsymbol{\theta}^Tf_1(\textbf{A})\boldsymbol{\theta}+\boldsymbol{\theta}^Tf_2(\textbf{B}),\label{eqn22}
\end{equation}
where $f_1:\mathbb{R}^{K\times K}\to\mathbb{R}^{K\times K}$ and $f_2:\mathbb{R}^{K\times K}\to\mathbb{R}^{K}$ are the linear maps, which can be defined as
\begin{subequations}
\begin{equation}
[f_1(\textbf{A})]_{i,j}=
\begin{cases}
\textbf{A}_{i.j}-\sum\limits_{n=1}^K\textbf{A}_{i,n}, & \mbox{if }i=j,\\
\textbf{A}_{i.j}, & \text{otherwise}.
\end{cases},\label{eqn23}
\end{equation}
and
\begin{equation}
[f_2(\textbf{B})]_i=2\sum_{n=1}^K\textbf{B}_{i,n}, \label{eqn24}
\end{equation}
\end{subequations}
The proof of Lemma 1 can be referred to \cite{lemma1}. Based on Lemma 1, we can obtain an approximation expression of the outage constraint (\ref{op31}). Furthermore, if $\sigma_{m,n}$ is small, we have:
\begin{equation}
\begin{aligned}
&\text{Pr}\left\{\textbf{q}_{m,n}^H\textbf{Z}_{m,n}\textbf{q}_{m,n}\leq\sigma^2_0\right\}\approx\\
&\text{Pr}\left\{\sum\limits_{i,j}\textbf{Z}_{m,n,[i,j]}+\boldsymbol{\nu}^T\textbf{Q}_{m,n}\boldsymbol{\nu}+2\boldsymbol{\nu}^T\textbf{r}_{m,n}\leq\sigma^2_0\right\}\leq p_{m,n},\\ \label{eqn25}
\end{aligned}
\end{equation}
where $\boldsymbol{\nu}\sim\mathcal{N}(\textbf{0},\textbf{I})$ is a $K$-dimentional Gaussian random vector, $\textbf{Q}_{m,n}=\sigma^2_{m,n}\textbf{C}^{\frac{1}{2}}_{m,n}f_1(\textbf{A}_{m,n})\textbf{C}^{\frac{1}{2}}_{m,n}$, and $\textbf{r}_{m,n}=\frac{1}{2}\sigma_{m,n}\textbf{C}^{\frac{1}{2}}_{m,n}f_2(\textbf{B}_{m,n})$. Then, we derive the equivalent convex restrictions of (\ref{eqn25}), which can be obtained by using the following lemma:

\emph{Lemma 2:} Let $\textbf{e}\sim\mathcal{N}(\textbf{0},\textbf{I})$, $\textbf{Q}\in\mathbb{H}^{K\times K}$ and $\textbf{r}\in\mathbb{R}^{K\times K}$ being known. For any $\mu>\frac{1}{\sqrt{2}}$ and $\tau>0$, we have
\begin{equation}
\begin{aligned}
&\text{Pr}\left\{\textbf{e}^T\textbf{Q}\textbf{e}+2\text{Re}\left\{\textbf{e}^T\textbf{r}\right\}+s\leq0\right\}\\
&\leq
\begin{cases}
\exp\left(-\frac{\tau^2}{4T^2}\right)  &0<\tau\leq2\lambda\mu T,\\
\exp\left(-\frac{\tau\lambda\mu}{T}+(\lambda\mu)^2\right)  &\tau>2\lambda\mu T,
\end{cases}\label{eqn26}
\end{aligned}
\end{equation}
where $s=\tau-\text{tr}(\textbf{Q})$, $\lambda=1-\frac{1}{2\mu^2}$, and $T=\mu\|\textbf{Q}\|_F+\frac{1}{\sqrt{2}}\|\textbf{r}\|$. The proof of Lemma 2 can be referred to \cite{lemma2}. Note that if we find a proper $\tau$, we can make Lemma 2 approximately equal to constraint (\ref{eqn25}). Thus, we let the right-side expressions of the (\ref{eqn26}) be equal to $p_{m,n}$, respectively, and thus get the following equivalent equations:
\begin{subequations}
\begin{equation}
\tau_1=2\sqrt{\ln{\frac{1}{p_{m,n}}}}T, \label{eqn27}
\end{equation}
\begin{equation}
\tau_2=(\lambda\mu+\frac{\ln{\frac{1}{p_{m,n}}}}{\lambda\mu})T, \label{eqn28}
\end{equation}
\end{subequations}
In particular, it is observed that the right side of (\ref{eqn26}) are all monotonous decreasing functions about $\tau$. Therefore, we only need to find the minimum value of $\tau_1$ and $\tau_2$. It is obvious that the minimum value of $\tau_2$ is equal to $\tau_1$ if $\lambda\mu=\sqrt{\ln\frac{1}{p_{m,n}}}$. The parameter $\mu$, which satisfies $\mu>\frac{1}{\sqrt{2}}$, can be obtained according to the definition of $\lambda$. In conclusion, we can get the least conservative approximation of (\ref{eqn25}), which is given by
\begin{equation}
\tau\geq2\sqrt{\ln{\frac{1}{p_{m,n}}}}T. \label{eqn29}
\end{equation}
Combining with the definition of $T$, (\ref{eqn29}) can be replaced by a group of second-order cone (SOC) constraints as follows:
\begin{subequations}
\begin{equation}
\text{tr}(\textbf{Q}_{m,n})+s_{m,n}\geq2\sqrt{\ln{\frac{1}{p_{m,n}}}}(x_{m,n}+y_{m,n}), \label{eqn30}
\end{equation}
\begin{equation}
\frac{1}{\sqrt{2}}\|\textbf{r}_{m,n}\|\leq x_{m,n}, \label{eqn31}
\end{equation}
\begin{equation}
\mu_{m,n}\|\textbf{Q}_{m,n}\|_{F}\leq y_{m,n}, \label{eqn32}
\end{equation}
\end{subequations}
where $s_{m,n}=\sum\limits_{i,j}\textbf{Z}_{m,n,[i,j]}-\sigma^2_0$. In addition, similar to Algorithm 1, we replace the rank-one constraint by a penalty function. Finally, problem (\ref{OP3}) can be reformulated as (\ref{OP4}) at the top of the next page, where $\rho_2$ is the penalty factor.
\newcounter{TempEqCnt2}  
\setcounter{TempEqCnt2}{35} 
\setcounter{equation}{34}
\begin{figure*}[ht]
\begin{equation} \label{OP4}
\min\limits_{\boldsymbol{\mathbf{W}}_m, \forall m, x_{m,n},y_{m,n}, \forall m,n}\sum_{m=1}^{M}\text{tr}(\textbf{W}_m)+\rho_2\sum\limits_{m=1}^M\left(\text{tr}\left(\textbf{W}^{(t+1)}_m\right)-\left(\textbf{v}^{(t)}_{m,max}\right)^H\textbf{W}^{(t+1)}_m\textbf{v}^{(t)}_{m,max}\right),
\end{equation}
\begin{equation}
\textrm{s.t.}\:(\ref{op32}),(\ref{op33}),(\ref{eqn30}),(\ref{eqn31}),(\ref{eqn32}). \nonumber
\end{equation}
\hrulefill
\end{figure*}
\setcounter{equation}{\value{TempEqCnt2}}
Thus, problem (\ref{OP4}) reduces to a SDP problem, which can be effectively solved by CVX. In summary, the critical robust design can be described as Algorithm 2.

\begin{algorithm}
\caption{: Critical Robust Design of Multi-beam LEO Satellite IoT for the Total Power Consumption Minimization}
\label{alg2}
\hspace*{0.02in} {\bf Input:}
$K,M,N_m,\alpha_{m,n},\gamma_{m,n},\mu_{m,n},\sigma_0^2,P_k$ and $p_{m,n}$.\\
\hspace*{0.02in} {\bf Output:}
$\textbf{w}_m$.
\begin{algorithmic}[1]
\STATE{\textbf{Initialize} $ x_{m,n}=0,y_{m,n}=0$ and feasible solution $\textbf{W}^{(0)}_m$ by solving problem (\ref{OP3}) dropping the constraint (\ref{op34}); }
\STATE{\textbf{Set} accuracy $\epsilon_2$, maximal iteration number $T_{\max}$, penalty factor $\rho_2$ and coefficient $\kappa_2$.}
\STATE{\textbf{repeat}}
\STATE{\quad\textbf{Generate} $\textbf{Z}_{m,n}=\textbf{A}_{m,n}+j\textbf{B}_{m,n}$, and calculate $f_1(\textbf{A}_{m,n})$, $f_2(\textbf{B}_{m,n})$;}
\STATE{\quad solve problem (\ref{OP2}) by CVX, then obtain $\textbf{W}_m^{(t)}$.}
\STATE{\quad\textbf{if} $\left|\text{tr}\left(\textbf{W}^{(t)}_m\right)-\lambda^{(t)}_{m,\max}\right|>\epsilon_2$ \textbf{then}}
\STATE{\qquad update the penalty factor $\rho^{(t)}_1=\kappa_2\rho^{(t+1)}_1$.}
\STATE{\quad\textbf{end if}}
\STATE{\quad Update iteration number $t=t+1$.}
\STATE{\textbf{until} $t=T_{\max}$ or solution converges.}
\STATE{\textbf{Finally}, use eigenvalue decomposition (EVD) to $\textbf{W}_m^{(t)}$ and obtain $\textbf{w}^*_{m}$ according to (\ref{new3}).}
\end{algorithmic}
\end{algorithm}

\subsection{Complexity Analysis}
\begin{figure}[t] \centering
\includegraphics [width=0.48\textwidth] {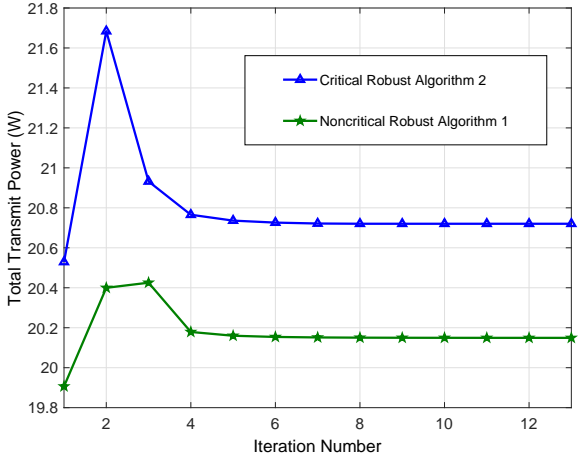}
\caption {Convergence behavior of the proposed Algorithm 1 and Algorithm 2.}
\label{Figcov}
\end{figure}

In this part, we analyze the computational complexity of the two proposed robust algorithms. Obviously, the problems (\ref{OP2}) and (\ref{OP4}) only involve linear matrix inequality (LMI) and second-order cone constraints. Thus, we can utilize the standard interior-point method (IPM) to investigate the computational complexity of the both algorithms. According to generic IPM \cite{IPM}, the complexity often consists of two parts, namely, iteration complexity and iteration computation cost. For a certain $\zeta>0$, the iteration complexity of an $\zeta-$optimal solution is in the order of $\Psi\ln{\frac{1}{\zeta}}$, where $\Psi$ is the barrier parameter evaluating the geometric complexity of conic constraints. Moreover, the computation cost is decided by construction and factorization of coefficient matrix. It is assumed that decision variables in problems (\ref{OP2}) and (\ref{OP4}) are real-valued \cite{complex}. Note that problem (\ref{OP2}) has $M$ LMI constraints of size 1 and $2M$ LMI constraints of size $K$, problem (\ref{OP4}) has $M$ LMI constraints of size 1, $2M$ LMI constraints of size $K$ and $2M$ SOC constraints of size $K+1$. We summarize the computational complexity of the two proposed algorithms in Table \ref{TAB1} on the next page. Furthermore, it is seen from Fig. \ref{Figcov} that the both proposed algorithms can be converged after no more than 8 times iterations as long as we set proper penalty factors. Thus, the two proposed algorithms have low computational complexity for practical deployment.
\begin{table*}[t]
\caption{MAIN PARAMETERS ON LEO SATELLITE}
\label{TAB1}
\centering
\begin{spacing}{1.2}
\begin{tabular}{|c|c|}
\hline
Algorithms & Complexity in order of $\ln{\frac{1}{\zeta}}\Upsilon$ with $n=\mathcal{O}(MK^2)$\\
\hline
Algorithm 1 & $\Upsilon=\sqrt{M(2K+1)}\cdot n\cdot[M(n+1)+KM(1+n+K^2+nK)+n^2]$ \\
\hline
Algorithm 2 & $\Upsilon=\sqrt{M(2K+5)}\cdot n\cdot [M(K+1)(2K+n+3)+MK^2(K+n)+n^2]$\\
\hline
\end{tabular}
\end{spacing}
\end{table*}

\section{Numerical Results}
In this section, we provide extensive simulation results to evaluate the performance of the two proposed algorithms. The setup of simulation parameters is given in Table \ref{TAB2}.

\begin{table}[t]
\caption{MAIN PARAMETERS ON LEO SATELLITE}
\label{TAB2}
\centering
\begin{spacing}{0.95}
\begin{tabular}{|c|c|}
\hline
Parameter & Value\\
\hline
Orbit & LEO\\
\hline
Number of beams & 10\\
\hline
Number of total users/intra-region user & 30/3\\
\hline
Number of satellite antenna feeds & 60\\
\hline
Imperfect SIC coefficient & 0.05\\
\hline
Bandwidth & 25 MHz\\
\hline
Variance of AWGN & 1\\
\hline
Carrier frequency & 20 GHz\\
\hline
Altitude of orbit & 1000 km\\
\hline
Satellite antenna gain & 17 dBi\\
\hline
Receiver gain to noise temperature & 34 dB/K\\
\hline
Boltzmann's constant & $1.38 \times 10^{-23}$ J/m\\
\hline
Rain fading mean & -2.6 dB\\
\hline
Rain fading variance & 1.63 dB\\
\hline
3dB Angle & $0.4^\circ$\\
\hline
Iterative accuracy & $10^{-10}$\\
\hline
Variance of phase error & $5^\circ$\\
\hline
Normalized covariance matric & $\textbf{I}$\\
\hline
\end{tabular}
\end{spacing}
\end{table}

\begin{figure}[t] \centering
\includegraphics [width=0.5\textwidth] {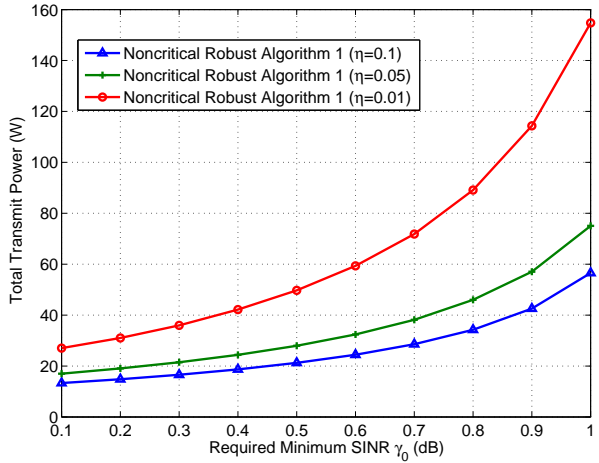}
\caption {Total transmit power versus required minimum SINR for different imperfect SIC coefficients on noncritical robust Algorithm 1.}
\label{Fig2}
\end{figure}

\begin{figure}[t] \centering
\includegraphics [width=0.5\textwidth] {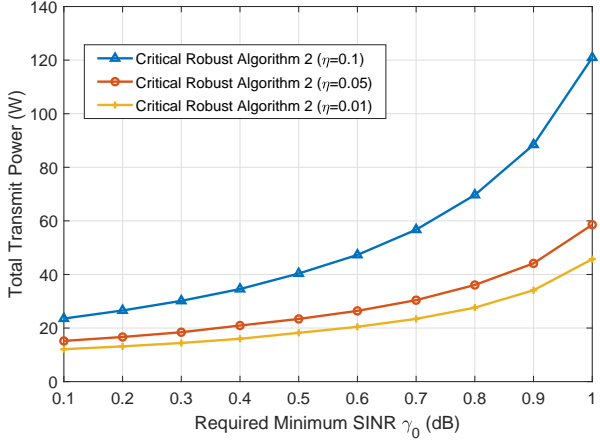}
\caption {Total transmit power versus required minimum SINR for different phase uncertainty on noncritical robust Algorithm 2.}
\label{Fig3}
\end{figure}

First, we reveal the impact of imperfect SIC coefficients on noncritical robust Algorithm 1 and critical robust Algorithm 2. From Fig. \ref{Fig2} and Fig. \ref{Fig3}, it is seen that the performance gaps on different imperfect SIC coefficients are subtle when the SINR threshold is small, which means the two proposed algorithms can both resist the influence of imperfect SCI in range of low SINR. However, as the SINR requirement increases, the growth rate of the total power consumption under the condition of large imperfect SIC coefficient ($\eta=0.1$) is much higher than that of small coefficients ($\eta=0.05$ or $\eta=0.01$). Due to the impact of imperfect SIC, power needed by UEs in the latter of efficient channel order to fight against intra-region interference is in direct proportion to imperfect SIC coefficients. Thus, the performance of SIC is a key factor to both Algorithm 1 and Algorithm 2.

\begin{figure}[t] \centering
\includegraphics [width=0.5\textwidth] {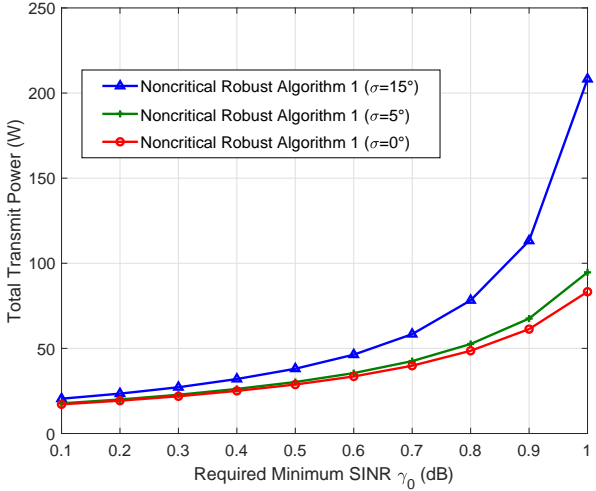}
\caption {Total transmit power versus required minimum SINR for different phase uncertainty on noncritical robust Algorithm 1.}
\label{Fig4}
\end{figure}

\begin{figure}[t] \centering
\includegraphics [width=0.5\textwidth] {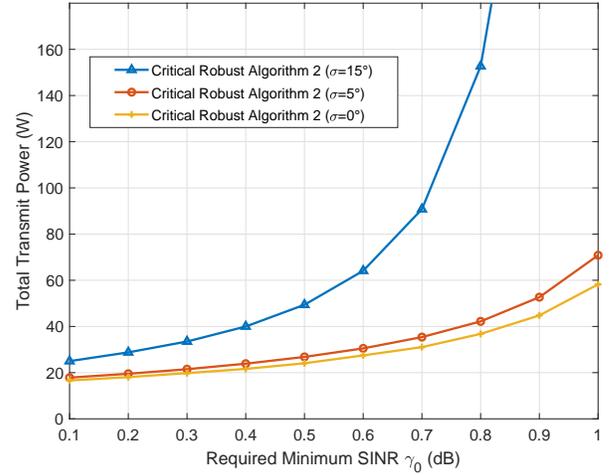}
\caption {Total transmit power versus required minimum SINR for different phase uncertainty on critical robust Algorithm 2.}
\label{Fig5}
\end{figure}
Second, we investigate the capability of combating channel phase uncertainty of the two proposed algorithms. The simulation trends of Fig. \ref{Fig4} and Fig. \ref{Fig5} are consistent with theoretical analysis, that is, the power consumption increases with the raise of phase uncertainty level. Besides, it is seen that the two proposed algorithms have a good robustness because of the small gap between the cases of $\sigma=0^\circ$ and $\sigma=5^\circ$. But with the increase of the phase error, Algorithm 2 performs worse than Algorithm 1. This is because the average SINR constraint is slacker than the outage probability constraint, which means that Algorithm 2 can meet a higher quality of service (QoS) requirement.

\begin{figure}[t] \centering
\includegraphics [width=0.51\textwidth] {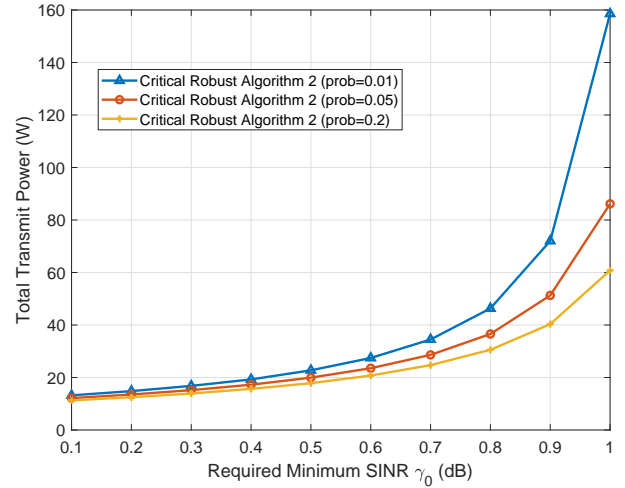}
\caption {Total transmit power versus required minimum SINR for different outage probability on critical robust Algorithm 2.}
\label{Fig6}
\end{figure}

Fig. \ref{Fig6} depicts the total transmit power consumption versus ruquired minimum SINR for different outage probabilities on Algorithm 2. As seen in Fig. \ref{Fig6}, the gap between the cases of $prob=0.01$ and $prob=0.05$ is larger than that between $prob=0.05$ and $prob=0.2$. We can deduce that with the raise of outage probability, the total transmit power consumption will decrease slower, which means Algorithm 2 is sensitive to the outage probability in a small range. In fact, outage probability can not be too large due to the QoS requirement in practice.

\begin{figure}[t] \centering
\includegraphics [width=0.51\textwidth] {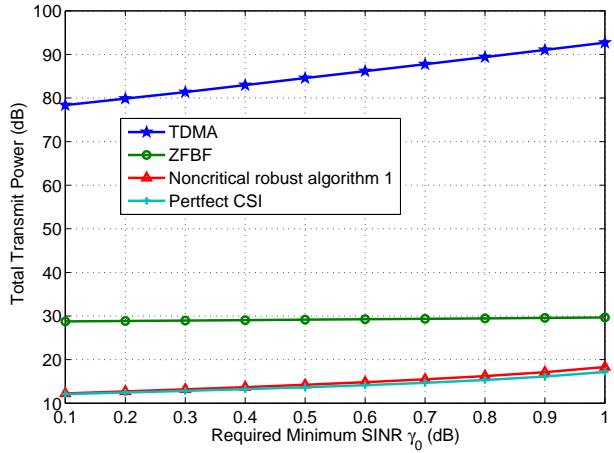}
\caption {The performance comparison of different algorithms under average SINR constraint.}
\label{Fig7}
\end{figure}

\begin{figure}[t] \centering
\includegraphics [width=0.5\textwidth] {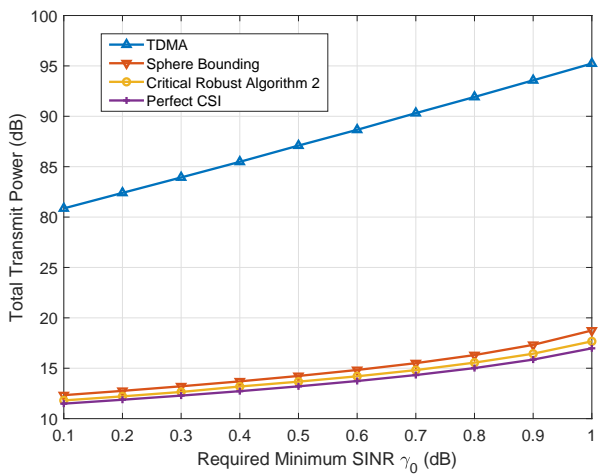}
\caption {The performance comparison of different algorithms under outage probability constraint.}
\label{Fig8}
\end{figure}

\begin{figure}[t] \centering
\includegraphics [width=0.5\textwidth] {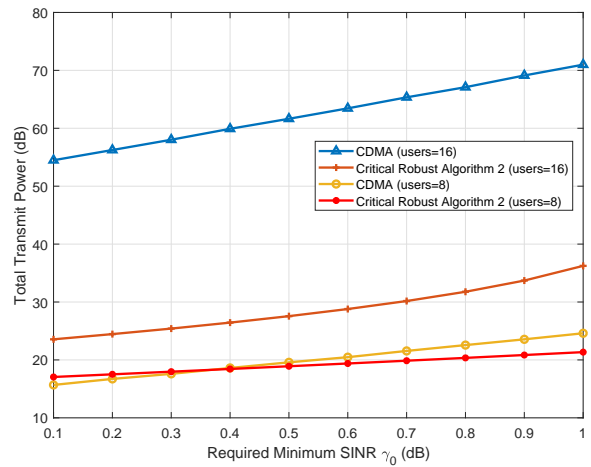}
\caption {The comparison of total transmit power required SINR threshold between CDMA and critical robust design.}
\label{Fig11}
\end{figure}

In Fig. \ref{Fig7} and Fig. \ref{Fig8}, we compare the performance of different algorithms under the same conditions. In addition, the coordinate of Y-axis changes into dB form and Fig.\ref{Fig8} omits the range of 25dB to 75 dB due to the large power comsumptation of TDMA. From Fig. \ref{Fig7}, it is obvious that TDMA consumes highest power. Although there is no interference between UEs in OMA schemes, each UE needs much higher transmit power to satisfy the SINR constraint compared with the NOMA scheme. Also the proposed Algorithm 1 has a better performance than the zero-forcing beamforming (ZFBF) method. ZFBF has a strong advantage of alleviating the inter-beam interference. However, in the range of lower SINR, the inter-beam interference is small such that the performance is worse than proposed Algorithm 1. In Fig. \ref{Fig8}, TDMA also performs worst among the all methods. The proposed Algorithm 2 has lower power consumption than S-Bounding method \cite{S-B} which proves the effectiveness of Algorithm 2. Besides, the difference of performance between two proposed algorithms and perfect CSI are subtle which demonstrates the powerful robustness of two algorithms.

\begin{figure}[t] \centering
\includegraphics [width=0.5\textwidth] {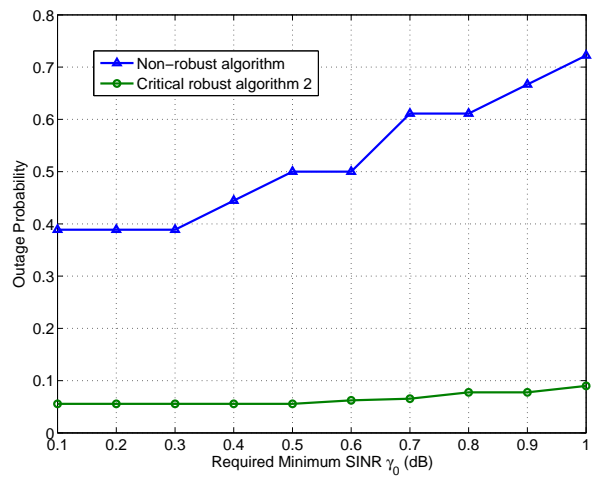}
\caption {The comparison of actual probability versus required SINR threshold between non-robust design and critical robust design.}
\label{Fig10}
\end{figure}

In Fig. \ref{Fig11}, we compare the total power consumption of NOMA and CDMA (we assume that spread spectrum code has good orthogonality and consider the process gain with different chip rate \cite{OCDMA} \cite{CDMA4}) under the same quality of service (QoS) requirement and the same spectrum efficiency. It is seen from Fig. \ref{Fig11} that CDMA slightly outperforms NOMA in low SINR requirement with a small number of users. However, with the increase of SINR requirement, NOMA has a better performance than CDMA. In the case of a large number of users, NOMA consumes much lower transmit power than CDMA. In other words, under the same condition of power resource and frequency resource, the NOMA-based satellite network can support much more devices than the CDMA-based one, which proves the effectiveness of the proposed NOMA scheme in LEO satellite IoT networks.

In Fig. \ref{Fig10}, the outage performance of the critical robust algorithm and non-robust scheme are depicted. In non-robust scheme, we consider the perfect CSI in the beamforming design, which leads to a serious mismatch between the designed beams and the actual channels. With considering the phase error into robust design, it is obvious that our proposed robust designs can guarantee the QoS of users and provide significant reduction in outage probability than the non-robust scheme. Thus, the simulation results prove the robustness and effectiveness of our proposed robust algorithms.

\section{Conclusion}
In this paper, we have designed a massive access framework for LEO multi-beam satellite IoT networks. To minimize the total power consumption of the LEO satellite under practical but adverse conditon of channel phase uncertainty, we propose two robust beamforming algorithms, one for noncritical IoT applications, the other for critical IoT applications. Finally, extensive simulation results validated the effectiveness and robustness of the proposed algorithms.

\end{document}